# Single-pulse all-optical switching in amorphous $Dy_xCo_{1-x}$ and $Tb_xCo_{1-x}$

Zexiang Hu*, Jean Besbas, Ross Smith, Niclas Teichert, Gwenael Atcheson, Karsten Rode, Plamen Stamenov and J. M. D. Coey

*CRANN, AMBER and School of Physics, Trinity College Dublin, Dublin 2, Ireland.*

**Abstract**

Repeated uniform switching of the magnetization of thin films of ferrimagnetic amorphous $Gd_x(FeCo)_{1-x}$ in response to single fast laser pulses is well established. Here we report unusual toggle switching in thin films of sperimagnetic amorphous $Dy_xCo_{1-x}$ and $Tb_xCo_{1-x}$ with x ≈ 0.25 irradiated with single 200 fs pulses of 800 nm laser light. The samples have strong local random anisotropy due to the non-S state rare earth. The compensation temperature of the films is ≤ 180 K and their Curie temperature is ≈ 500 K. They are mostly switched by the first pulse, and subsequent pulses lead to partial re-switching of a decreasing amount of the irradiated area, with a granular structure of submicron regions of switched and unswitched material. Individual switched domains about 700 nm in size are observed around the edge of the irradiated spots where the fluence is at the threshold for switching. Results are discussed in terms of a random anisotropy model where the ratio of local anisotropy to exchange is temperature dependent and close to the threshold for strong pinning.

*zehu@tcd.ie

## INTRODUCTION

Manipulation of magnetization in thin films without applying a magnetic field is of great interest in spin electronics. Conventional methods of switching the magnetization by using an external directional stimulus such as a pulsed magnetic field or a spin-polarized current are limited to timescales of about 100 ps.[1] Single ultrafast laser pulses with durations of order 100 fs were originally shown by Beaurepaire *et al.* to *demagnetize* Nickel in about a picosecond.[2] Subsequently, single-pulse all-optical thermal *switching* of the magnetization (SP-AOS) was established by Ostler *et al.* in ferrimagnetic amorphous Gd-Fe-Co on a picosecond timescale,[3] and its observation was extended to other Gd-based thin film systems.[4,5] The first non-Gd based ferrimagnet to exhibit the effect was crystalline Mn-Ru-Ga,[6,7] where Banerjee *et al.* have shown that re-switching is possible after 10 ps.[8]

Radu *et al.*'s 2011 study of the dynamics of the Gd and Fe sublattices during spin reversal in their a-Gd-Fe-Co thin films after a 50 fs laser pulse found that an unexpected transient parallel alignment of the moments preceded the switching.[9] The same group confirmed that SP-AOS was essentially a thermal effect.[3] The magnetization could be repeatedly toggled back and forth between two orientations, and the effect was independent of the helicity of the light.

A general thermodynamical framework for this ultrafast laser-induced spin dynamics in multisublattice magnets was proposed by Mentink *et al.*.[10] A temperature-dominated regime yields distinct sublattice demagnetizing dynamics because of the different magnetic moments, with the Fe sublattice reaching zero first.[9] On the ps timescale, exchange relaxation is dominant and brings about the ferromagnetic-like state by transferring moment from the Gd to the fully demagnetized Fe. The magnetization of the Fe increases in the direction opposite to its initial as it cools one while the Gd is still decreasing, which finally leads to complete switching of the magnetization.

The phenomenon of helicity-dependent optical switching requiring many short pulses is quite different in origin; it was discovered by Stanciu *et al.* in 2007[11] and explored in a wide range of multilayer and alloy magnetic materials.[12,13]

Gd is a spherically symmetric ion and a-GdFeCo is ferrimagnetic due to the negative rare-earth (R) - transition metal (T) exchange interaction. However, for the non S-state R-T alloys such as Tb-Co or Dy-Co, the random local electrostatic fields acting on the non-spherical 4*f* charge distributions create random local anisotropy that influences the orientation of the R moments in the magnetically-ordered state. The contribution of R to the net ferrimagnetic magnetization below the compensation temperature is much reduced in a sperimagnetic ground state where the Co subnetwork is essentially collinear and the Dy or Tb subnetwork moments are distributed at random in a cone whose axis is antiparallel the cobalt.[14]

After SP-AOS was discovered in a-GdFeCo thin films, much work was done to look for the effect in systems with a different rare earth. This met with little success. Both a-Tb-Co[12,16,17] and a-Gd-Fe-Co[12] thin films were found to exhibit multi-pulse helicity-dependent switching at for compositions around 25 at%, which exhibit a compensation point where the net R and T subnetwork moments cancel. Film thickness and the size of the laser spot size relative to domain size were criteria for observing the effect.[16] SP-AOS is observed in a series of $(Gd_{22-x}Tb_x)Co_{78}$ alloys for all except the x = 22 Tb end member[18]. Transient sub-picosecond reversal has been seen in a study of the demagnetization dynamics of $Tb_{26}Co_{74}$.[19] Single circularly-polarized 150 fs laser pulses have been found to create skyrmions as small as 150 nm in diameter in $Tb_{22}Co_{60}Fe_9$.[20] A study by Liu *et al.*[21] used gold two-wire nanoantennas placed on a-$Tb_{17}Fe_{72}Co_{11}$ thin films for near-field enhancement of the laser pulse to reversibly switch single 50 nm domains in the film. A recent report describes helicity-independent toggle switching of Tb/Co electrodes in a magnetic tunnel junction, but not of the amorphous alloy.[22]

Based on these results, we have examined a-TbCo$_x$ or a-DyCo$_x$ alloys for evidence of SP-AOS and signs of the influence of local random anisotropy, with the view to obtain a better understanding of the SP-AOS phenomenon.

**EXPERIMENTAL METHODS and RESULTS**

All thin films were deposited on Si wafer substrates with 500 nm thermal oxide at room temperature by DC magnetron sputtering. The a-DyCo thin layers were deposited in a Shamrock system with a base pressure of 2 x 10$^{-8}$ Torr. The a-TbCo thin films were grown using the ultra-high vacuum multichamber deposition and characterization tool Trifolium Dubium. They were co-sputtered from rare-earth (Dy or Tb) and Co targets. Composition was controlled by changing the RE target plasma power while fixing the Co power. The samples were capped with a protective layer of either SiO$_2$ or Al$_2$O$_3$. A superconducting quantum interference device (SQUID) was used to measure the magnetization of the specimens. Both films studied here exhibit perpendicular anisotropy. Hysteresis was measured in an Evico MOKE microscope and by anomalous Hall effect. Structural and magnetic information on two representative samples is summarized in Table I.

TABLE I. Characteristics of the samples. Magnetic values are at 296 K.

| Nominal Composition | Thickness (nm) | Capping layer | RMS roughness (nm) | $T_{comp}$ (K) | $M_s$ (kA/m) | $M_s$ ($\mu_B$/f.u.) | $\mu_0 H_c$ (mT) | Energy density (mJ/cm$^2$) |
|---|---|---|---|---|---|---|---|---|
| **Tb$_{0.25}$Co$_{0.75}$** | 20.0 | Al$_2$O$_3$ | 0.34 | ≪RT | 240 | 0.41 | 44.3 | 2.2 |
| **Dy$_{0.25}$Co$_{0.75}$** | 10.3 | SiO$_2$ | 0.36 | 180 | 175 | 0.29 | 35.0 | 0.8 |

The X-Ray diffraction measurements in Fig. 1 show only peaks associated with the silicon substrate, including a broad forbidden (002) peak due to stacking faults in the Si. There are no peaks associated with any other crystalline phase in these amorphous sputtered films.[23] Fitting small angle X-ray reflection gives values for the film thickness and density. The RMS surface roughness for both films, deduced from atomic force microscopy (AFM), is 0.34 nm.

Hysteresis loops of the films are presented in Fig. 2. The data reflect the cobalt magnetization. The RE-TM coupling is antiferromagnetic, but the ferrimagnetism is modulated by random anisotropy on the rare earth that increases the net magnetization above compensation. The magnetic ordering temperature deduced from extrapolation of the temperature dependence of the Dy sample is 500 K. The compensation temperature, where the net magnetization crosses zero and the coercivity diverges is 180 ± 3 K. for a-Dy$_{0.25}$Co$_{0.75}$. The Curie temperature of the Tb sample is expected to be similar, but its compensation temperature is much lower. The atomic moment on Co measured to be 1.5 μ_B from a ferromagnetic a-Y_0.25Co_0.75 film, hence the moments of Dy and Tb at room temperature are deduced as 3.4 μ_B and 2.9 μ_B, respectively. They are much reduced from the atomic values of 10 μ_B and 9 μ_B by random anisotropy and weak R-T exchange.

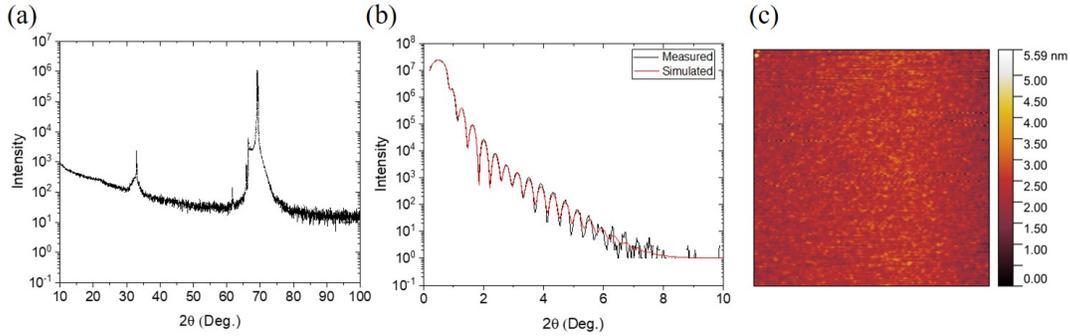

Fig. 1. (a) X-ray diffraction, (b) X-ray reflectivity and (c) AFM characterization of the Tb$_{0.25}$Co$_{0.75}$ film. The area in (c) is 5 × 5 μm.

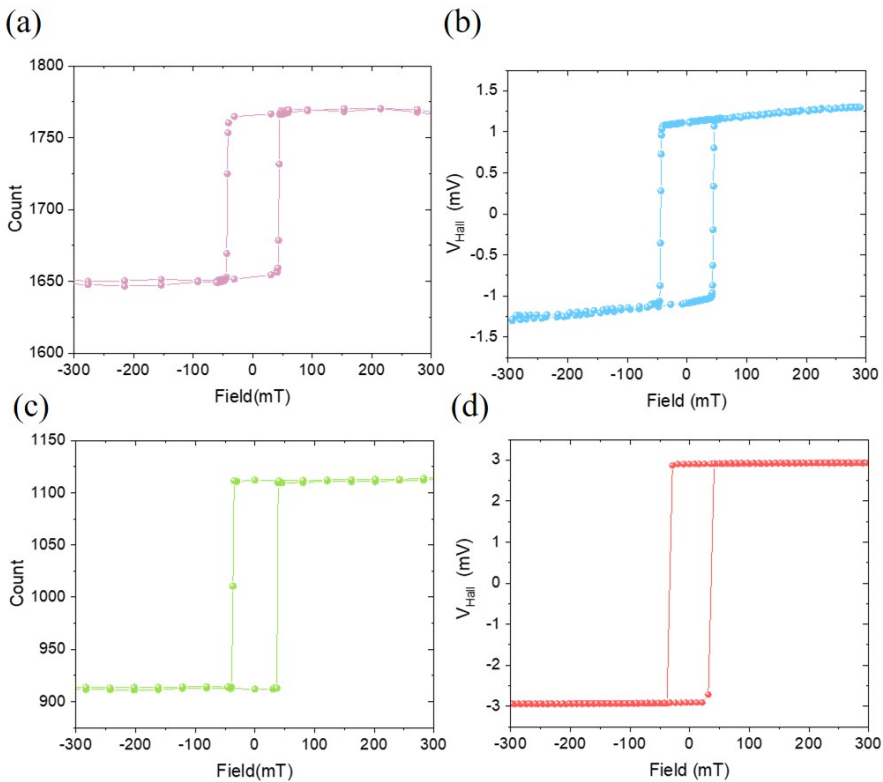

Fig. 2. Room-temperature hysteresis loops measured by anomalous Hall effect (left) and polar Kerr effect in blue light (right) for Tb$_{0.25}$Co$_{0.75}$ (top) and Dy$_{0.25}$Co$_{0.75}$ (bottom).

The effect of irradiation by single 200 fs laser pulses was investigated *in-situ* using a homemade polar MOKE microscope in the setup shown in Fig. 3(a). High-resolution images were measured *ex-situ* in the Kerr microscope. The 800 nm pulses were generated by a Ti:sapphire-based laser amplifier operating in single-pulse mode. Energy was controlled by rotating the half-waveplate (HWP), followed by a linear polarizer in the pulse beam path. Energy was calibrated by measuring the average beam power with the laser amplifier operating at 1 kHz. A parabolic mirror focused the $TEM_{00}$ beam with Gaussian intensity profile onto the sample surface down to a spot of diameter of 90 um, full width at half-maximum, calibrated using the Liu method.[24] Prior to irradiation, the net magnetization $M_{net}$ was saturated in a 0.5 T field directed outwards. For both samples the magnetization of the Co network is dominant at room temperature and initially in the outward direction. This state appears as a uniform grey area in the unirradiated films shown in Fig. 3(b). Irradiation by a single pulse at low fluence does not alter the magnetic state, but above a certain threshold, a durable effect of the light pulse is recorded as in irregular black patch of switched magnetization at the center of image as shown in Fig. 3(b), where the magnetization points inwards. The threshold fluence is lower for the 10 nm thick $DyCo_3$ than for 20 nm thick $TbCo_{3,4}$ as seen in Table I. The threshold fluence increases linearly with film thickness up to 40 nm. Switched areas seen after the first pulse in Figs 3 – 5, especially in the diffuse border region, are composed of domains 0.7 ± 0.3 µm in size. This behavior is quite different to single-pulse all-optical switching in similar conditions in amorphous $Gd_x(FeCo)_{1-x}$ or crystalline $Mn_2Ru_xGa$ films, where no such structure has been detected.[3,6]

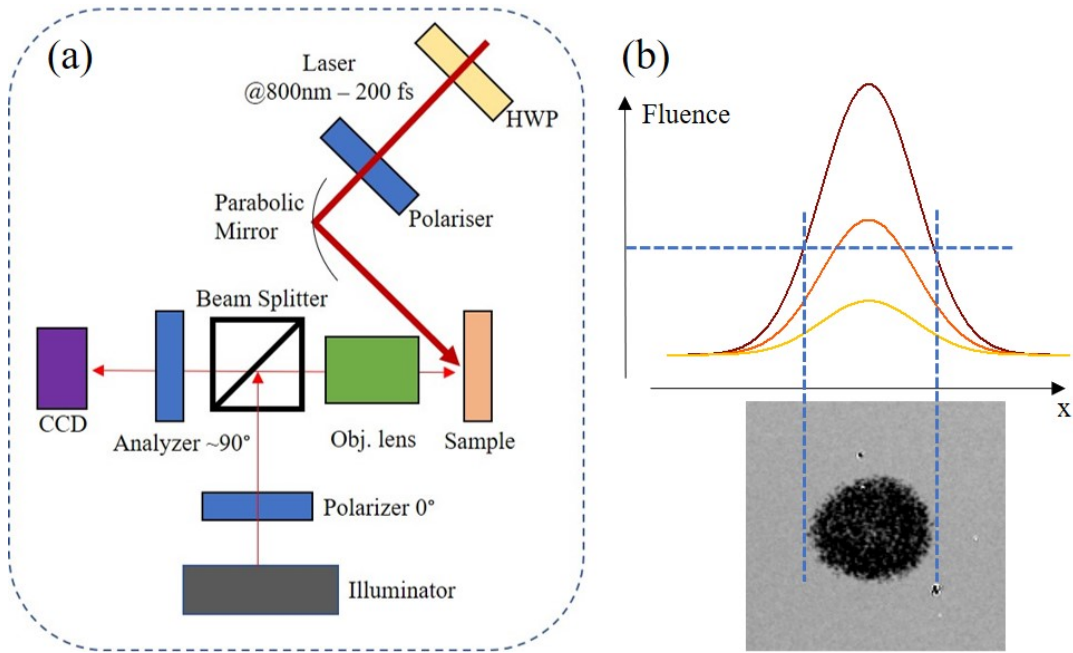

Figure 3: (a) Experimental setup using a parabolic mirror to focus a single 800 nm laser pulse onto the sample surface. (b) Gaussian profile of the fluence on the sample. Above a certain threshold fluence, the magnetisation of the $Dy_{0.25}Co_{0.75}$ sample after a single pulse reveals a mostly-reversed multidomain state.

We systematically irradiated regions of the samples with sequences of 1 to 10 pulses and examined them in the Evico microscope. Data are shown in Fig. 4 for sequences of 1 – 6 low- or medium- fluence pulses for each sample, separated by a 1 s delay. We focus our analysis here on the lower-fluence behavior, where rings are not evident. After an initial, nearly complete switch with the first pulse, the contrast between subsequent pulse pairs is reduced. The toggle switching was still present after six pulses but after 10 pulses it was much reduced.

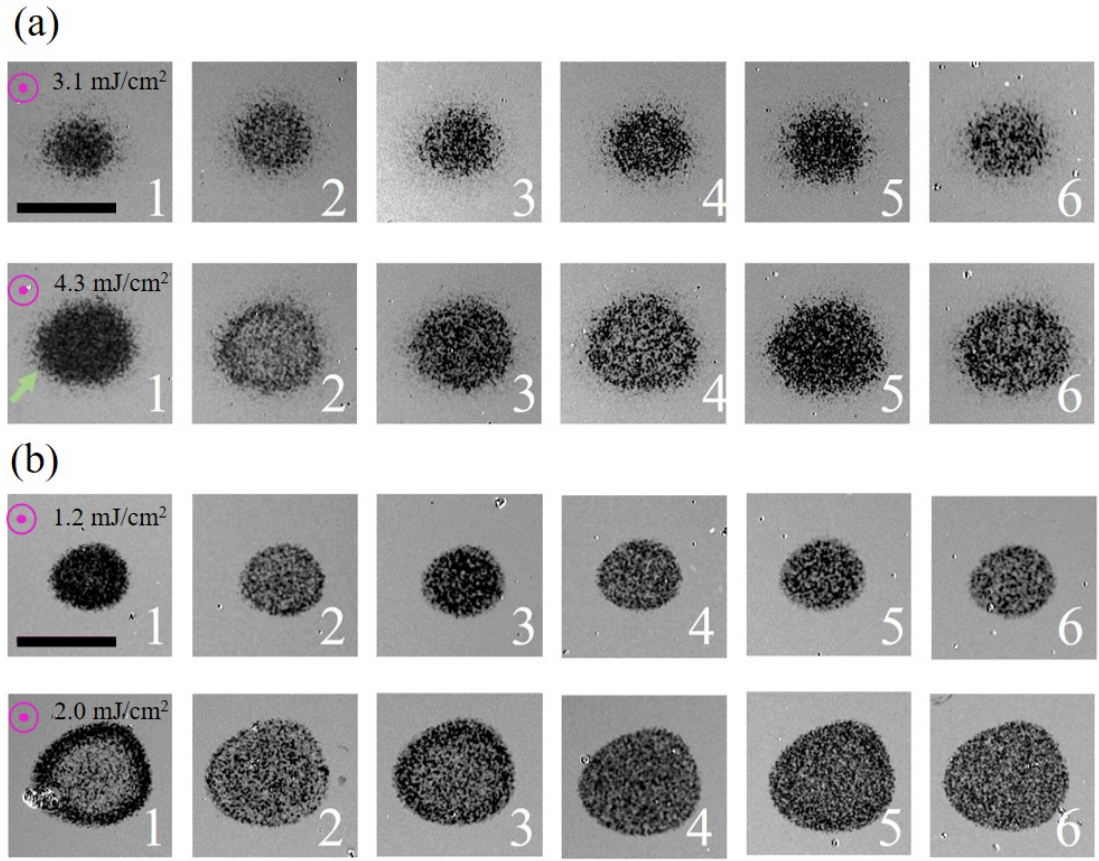

Fig. 4. Kerr images of films of (a) $Tb_{0.25}Co_{0.75}$ and (b) $Dy_{0.25}Co_{0.75}$ after a sequence of 200 fs laser pulses. The number of pulses is indicated on the figures. The fluences are noted at the beginning of each sequence. The scale bar is 50 μm.

Figure 5 presents a detailed analysis of higher-magnification (16 bit greyscale) images. The. background is flattened away from the region of interest using a biquadratic polynomial. The normalization, based on the removal of the initial outwardly magnetized state of the film, is accurate to within 5%. Elliptical regions of interest are fully centered within the irradiated spots. Histogram distributions of the local normalized data are constructed from the greyscale distributions within the elliptical regions of interest. Fig 5b) shows the distributions for the unperturbed film and after the first 1 – 3 shots; Fig. 5c) shows the peak of the magnetization distribution after each sequential shot. Error bars signify the positive and negative half widths at half maximum of the distributions. The first pulse reverses 75% of the domains, the next one 45 % and after 10 pulses, only 10 % is being reversed. The average size of the isolated domains is 0.7 ± 0.3 µm.

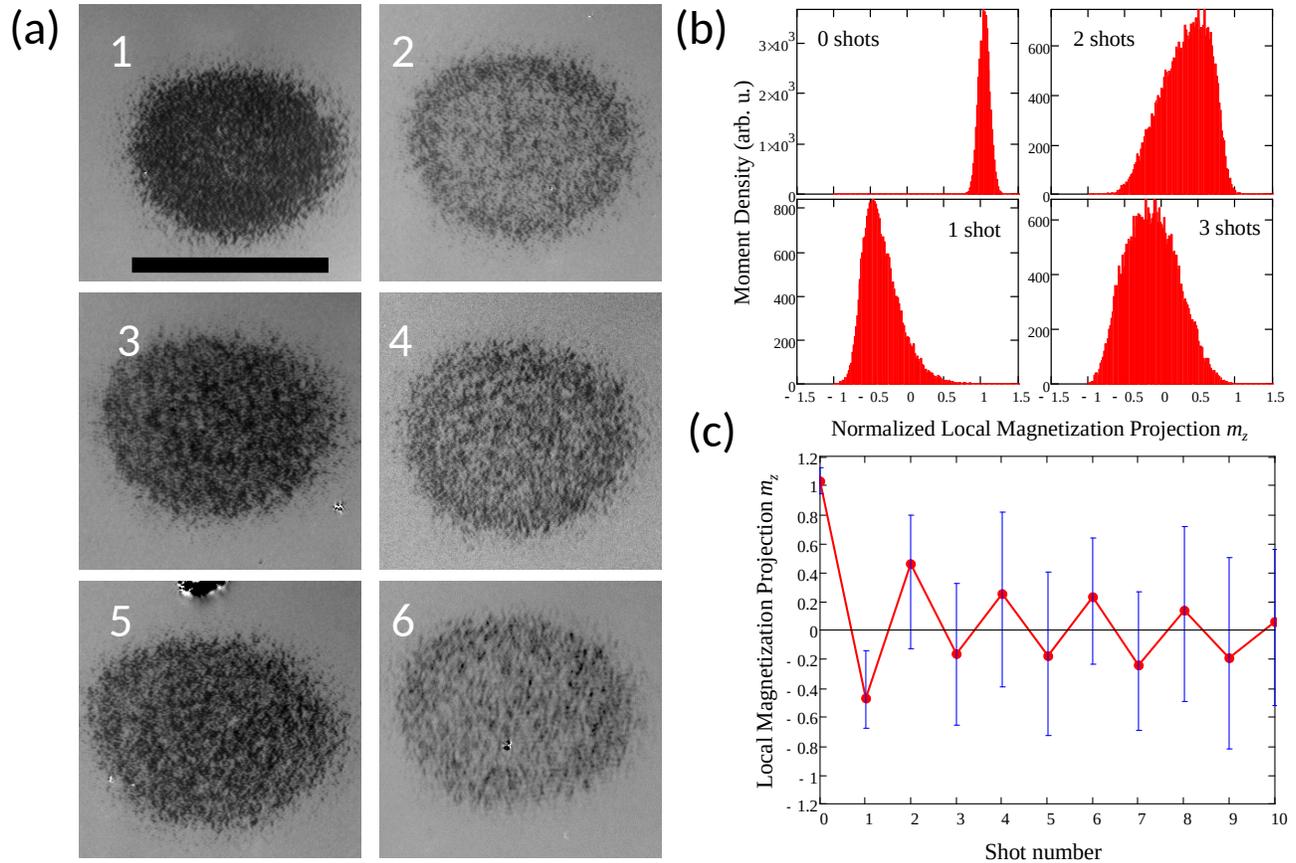

FIG. 5. (a) Kerr images of $Tb_{0.25}Co_{0.}$ films obtained after a sequence of 200 fs laser pulses. The scale bar is 50 µm. (b) The domain size distribution of the magnetized film, and after 0 - 3 pulses (c) The normalized fractions of outwardly or inwardly magnetized domains after 0 - 10 pulses.

**DISCUSSION**

The single-pulse all-optical toggle switching we observe in Figs 4 and 5 differs from that in $Gd_x(Fe,Co)_{1-x}$, $Mn_2Ru_xGa$ and anything previously reported for Dy-Co or Tb-Co. The difference is the nanoscale granularity of the switched areas, particularly evident around the borders of the spots where fluence is close to threshold. Only the study by Liu *et al.*, using nanoscale antennae reported thermal switching of a single domain, smaller than those we find here.[21]

We take the established mechanism for SP-AOS in a-$Gd_x(Fe,Co)_{1-x}$ as a basis for the discussion. The final step to form the switched state involves re-magnetization of the rare earth under the influence of antiferromagnetic exchange with the transition metal. With non-S-state rare earths, the process is modulated by the temperature-dependence of the growing local random anisotropy as well as the bulk perpendicular magnetic anisotropy of the film as it cools. The former establishes the 700 nm in-plane correlation length.[14, 15, 25]

Here the analytical[15] and computational[25] analysis of random local anisotropy in a ferromagnetcarried out in the 1970s provides some useful insight. Local random anisotropy in a ferromagnet is treated in the Harris-Plischke-Zuckermann model[15] where the first term is the interatomic exchange and the second is the locally random 'crystal field' interaction is

$$\mathcal{H}_{HPZ} = -2\sum_{i>j} \mathcal{J}_{i,j} \mathbf{S}_i \cdot \mathbf{S}_j - \sum_i D_i (\mathbf{e}_i \cdot \mathbf{S}_i)^2 \qquad (1)$$

Here $D_i$ is the local uniaxial 'crystal field' parameter, $\mathbf{e}_i$ is the local anisotropy direction, $\mathbf{S}_i$ is the total angular momentum of an atom and $\mathbf{e}_i \cdot \mathbf{S}_i$ is its z-component along a local z-axis. The key parameter is $\alpha$, the ratio $D/J$, the relative strength of random anisotropy versus exchange, which is independent of temperature in the one-sublattice random ferromagnet. There are two important results: One is that that computer simulations indicate a crossover from strong to weak local pinning of the magnetization by the anisotropy at $\alpha$ = 3.[25] The other is that the lengthscale *L* over which the magnetization axis varies is $L = (1/9\alpha^2)\pi^4 a$, where a is the average interatomic spacing.[15] In the present case the average distance between rare-earth ions deduced from the from the densities of the films is 0.5 nm. Taking *L* as the observed domain size, we find $\alpha \approx 0.1$. However, when applying the HPZ model to our amorphous ferrimagnets, $\alpha$ will *decrease* with increasing temperature because the random anisotropy term in the sum of $D_i(\mathbf{e}_i \cdot \mathbf{J}_i)^2$ and the exchange term varies roughly as $J_i$ because the exchange field from the cobalt is the main effect, and $S_z$ of Co, unlike $J_z$ of the rare earth varies little with temperature except just near $T_C$. At low temperature, in the sperimagnetic ground state, we estimate $\alpha > 1$; the pinning becomes strong.

Besides the local random anisotropy, the bulk perpendicular anisotropy around *x* = 0.25 is a well-established property of the amorphous R-Co films when R is a magnetic rare earth. Our finding that amorphous $Y_{0.25}Co_{0.75}$ films prepared in the same conditions are easy-plane suggests that the re-establishment of the rare earth moment is instrumental in establishing the perpendicular anisotropy needed to observer the switching and that it is *not* a property of the Co sublattice, which changes little during heating at the at the lower fluences. A possible distinguishing factor is the presence of low-lying excited *J*-multiplets for Dy or Tb that are absent for Gd.

**CONCLUSIONS**

The transient granular SP-AOS switching on a length scale of ~ 700 nm observed in $Dy_{25}Co_{75}$ and $Tb_{25}Co_{75}$ is quite distinct from that in $Gd_x(FeCo)_{1-x}$ where the toggle switching is sustained and uniform. Random atomic-scale anisotropy on the rare-earth subnetwork the sperimagnetic Dy and Tb alloys distinguishes them from their ferrimagnetic Gd counterpart.

The demagnetization rate of the high-moment atoms Dy and Tb is expected to be slower than that of Gd, and the re-magnetization process will be influenced by the local random anisotropy which increases faster than the exchange with decreasing temperature. The system is in the weak pinning region, consistent with ~ 700 nm independent granular switched regions.

Amorphous films containing rare earths with positive, negative and zero electric quadrupole moment and positive and zero magnetic moment provide an opportunity to disentangle the factors that influence the fast re-magnetization process of these interesting magnetic materials on both an atomic and a mesoscopic scale and extend the scope of picosecond-timescale investigations from local magnetization to local magnetic anisotropy.

**Acknowledgements**. This work was supported by Science Foundation Ireland under grants 16/IA/4534 ZEMS, 12/RC/2278 AMBER, 17/NSFC/5294 MANIAC and EU FET Open grant 737038 TRANSPIRE. Part of the research was carried out in the CRANN fast photonics laboratory, where we are grateful for technical help from Dr Jing Jing Wang. ZH acknowledges the support of a 1592 scholarship from Trinity College Dublin.